\documentclass[prl,twocolumn,twoside,preprintnumbers,superscriptaddress,nofootinbib]{revtex4}

\usepackage{amsmath,slashed}
\usepackage{graphicx,graphics,color}
\usepackage{dcolumn}
\usepackage[hyperfootnotes=false]{hyperref}
\usepackage{xspace}
\usepackage{enumerate}
\usepackage{varwidth}

\usepackage{physics}
\usepackage{amssymb}
\usepackage{mathtools}
\usepackage{dsfont}
\usepackage{multirow}

\hypersetup{
    colorlinks=true,
    linkcolor=blue,
    citecolor=blue
}

\newcommand{\mpi}{M_\pi}
\newcommand{\mpii}{M_{\pi^0}}

\newcommand{\beq}{\begin{equation}}
\newcommand{\eeq}{\end{equation}}

\newcommand{\Order}{\mathcal{O}}

\newcommand{\GeV}{\,\text{GeV}}

\newcommand{\Br}{\text{Br}}
\renewcommand{\Im}{\text{Im}\,}

\allowdisplaybreaks[1]

\begin{document}

\title{Improved calculation of radiative corrections to $\boldsymbol{\tau\to\pi\pi\nu_\tau}$ decays}

\author{Gilberto Colangelo}
\affiliation{Albert Einstein Center for Fundamental Physics, Institute for Theoretical Physics, University of Bern, Sidlerstrasse 5, 3012 Bern, Switzerland}
\author{Martina Cottini}
\affiliation{Albert Einstein Center for Fundamental Physics, Institute for Theoretical Physics, University of Bern, Sidlerstrasse 5, 3012 Bern, Switzerland}
\author{Martin Hoferichter}
\affiliation{Albert Einstein Center for Fundamental Physics, Institute for Theoretical Physics, University of Bern, Sidlerstrasse 5, 3012 Bern, Switzerland}
\author{Simon Holz}
\affiliation{Albert Einstein Center for Fundamental Physics, Institute for Theoretical Physics, University of Bern, Sidlerstrasse 5, 3012 Bern, Switzerland}

\begin{abstract} 
A reliable calculation of radiative corrections to $\tau\to\pi\pi\nu_\tau$ decays is an important prerequisite for using hadronic $\tau$ decays for a data-driven evaluation of the hadronic-vacuum-polarization contribution to the anomalous magnetic moment of the muon, $a_\mu^\text{HVP, LO}[\pi\pi,\tau]$. In this Letter, we present an improved model-independent analysis of these radiative corrections, including, for the first time, effects beyond point-like pions in the evaluation of the loop diagrams.
These structure-dependent corrections, implemented via a dispersive representation of the pion form factor, lead to significant changes compared to previous calculations due to enhancements near the $\rho(770)$ resonance. We also devise strategies for
the matching to chiral perturbation theory and
a stable implementation of the real corrections down to the two-pion threshold, which shows that some higher-order isospin-breaking corrections need to be kept due to a strong threshold enhancement. Finally, we perform dispersive fits to the currently available $\tau\to\pi\pi\nu_\tau$ spectra and discuss the consequences for isospin-breaking corrections in the evaluation of $a_\mu^\text{HVP, LO}[\pi\pi,\tau]$.
\end{abstract}

\maketitle

\emph{Introduction}---The release of the final results of the Fermilab experiment has significantly reduced the uncertainty of the experimental world average of the anomalous magnetic moment of the muon $a_\mu$~\cite{Muong-2:2025xyk,Muong-2:2023cdq,Muong-2:2024hpx,Muong-2:2021ojo,Muong-2:2021vma,Muong-2:2021ovs,Muong-2:2021xzz,Muong-2:2006rrc}
\beq
a_\mu^\text{exp}=11\,659\,207.15(1.45)\times 10^{-10} ,
\eeq
and would allow for a test of the Standard Model at an unprecedented level of precision.
Unfortunately, the limiting factor is represented by the precision of the current theory prediction~\cite{Aliberti:2025beg,Aoyama:2012wk,Hertzog:2025ssc,Volkov:2019phy,Volkov:2024yzc,Aoyama:2024aly,Parker:2018vye,Morel:2020dww,Fan:2022eto,Czarnecki:2002nt,Gnendiger:2013pva,Ludtke:2024ase,Hoferichter:2025yih,RBC:2018dos,Giusti:2019xct,Borsanyi:2020mff,Lehner:2020crt,Wang:2022lkq,Aubin:2022hgm,Ce:2022kxy,ExtendedTwistedMass:2022jpw,RBC:2023pvn,Kuberski:2024bcj,Boccaletti:2024guq,Spiegel:2024dec,RBC:2024fic,Djukanovic:2024cmq,ExtendedTwistedMass:2024nyi,MILC:2024ryz,FermilabLatticeHPQCD:2024ppc,Keshavarzi:2019abf,DiLuzio:2024sps,Kurz:2014wya,Colangelo:2015ama,Masjuan:2017tvw,Colangelo:2017qdm,Colangelo:2017fiz,Hoferichter:2018dmo,Hoferichter:2018kwz,Eichmann:2019tjk,Bijnens:2019ghy,Leutgeb:2019gbz,Cappiello:2019hwh,Masjuan:2020jsf,Bijnens:2020xnl,Bijnens:2021jqo,Danilkin:2021icn,Stamen:2022uqh,Leutgeb:2022lqw,Hoferichter:2023tgp,Hoferichter:2024fsj,Estrada:2024cfy,Deineka:2024mzt,Eichmann:2024glq,Bijnens:2024jgh,Hoferichter:2024vbu,Hoferichter:2024bae,Holz:2024lom,Holz:2024diw,Cappiello:2025fyf,Colangelo:2014qya,Blum:2019ugy,Chao:2021tvp,Chao:2022xzg,Blum:2023vlm,Fodor:2024jyn}
\beq
a_\mu^\text{SM}=11\,659\,203.3(6.2)\times 10^{-10},
\eeq
which is worse than the experimental one by about a factor of four. 
The main source of theory uncertainty is the leading-order (LO) hadronic-vacuum-polarization (HVP) contribution $a_\mu^\text{HVP, LO}$, which in Ref.~\cite{Aliberti:2025beg} was estimated on the basis of lattice-QCD calculations, differing substantially from the previous $e^+e^-$-based result~\cite{Aoyama:2020ynm,Davier:2017zfy,Keshavarzi:2018mgv,Colangelo:2018mtw,Hoferichter:2019mqg,Davier:2019can,Keshavarzi:2019abf,Hoid:2020xjs}. In fact, a data-driven estimate of this contribution was not even provided in Ref.~\cite{Aliberti:2025beg} due to tensions in the $e^+e^-\to \text{hadrons}$ data base. The main rationale for this change was that after the CMD-3 measurement of the critical $e^+e^-\to\pi^+\pi^-$ channel~\cite{CMD-3:2023alj,CMD-3:2023rfe} systematic tensions had increased to a level that could no longer be taken into account by a meaningful error inflation. Fortunately, there are a number of ongoing developments to rectify this situation~\cite{Aliberti:2025beg,Colangelo:2022jxc}, including new data (such as the preliminary measurement by BaBar~\cite{Polat:2026ysh}) as well as improvements of radiative corrections and Monte-Carlo generators~\cite{Campanario:2019mjh,Ignatov:2022iou,Colangelo:2022lzg,Monnard:2021pvm,Abbiendi:2022liz,BaBar:2023xiy,Budassi:2024whw,Aliberti:2024fpq,Fang:2025mhn,Budassi:2026lmr}.

A complementary avenue to shed light on the $2\pi$ contribution proceeds via $\tau\to\pi\pi\nu_\tau$ decays, in which case control over radiative corrections is arguably even more crucial. The $\tau$-based approach, pioneered in Ref.~\cite{Alemany:1997tn}, relies on an isospin rotation from the $\pi^\pm\pi^0$ channel probed in $\tau$ decays back to $\pi^+\pi^-$ as required for the two-pion contribution $a_\mu^\text{HVP, LO}[\pi\pi]$. To make this work, first, the $\tau$-specific radiative corrections need to be removed from the measured spectral function and branching fraction, then isospin-breaking (IB) corrections need to be applied to the hadronic matrix elements, to account for the difference between the pion form factors $F_\pi^V(s)$ and $f_+(s)$ in the neutral and charged channel, respectively, and finally the $e^+e^-$-specific IB corrections need to be added back to match the conventional definition of $a_\mu^\text{HVP, LO}[\pi\pi]$. The latter can be extracted in a reliable way using a dispersive approach~\cite{Colangelo:2018mtw,Colangelo:2020lcg,Colangelo:2022prz,Hoferichter:2023sli,Stoffer:2023gba,Leplumey:2025kvv} from the $e^+e^-\to\pi^+\pi^-$ data sets~\cite{Achasov:2006vp,CMD-2:2006gxt,BaBar:2012bdw,BESIII:2015equ,KLOE-2:2017fda,CMD-3:2023alj}, while the remaining IB corrections at present require model-dependent input~\cite{Aliberti:2025beg} (see Ref.~\cite{Colangelo:2025iuq} for a first step towards a dispersive calculation), preventing the use of $\tau$ decays in the evaluation of $a_\mu^\text{HVP, LO}[\pi\pi]$.

In this Letter, we address an improved calculation of the radiative corrections to $\tau\to\pi\pi\nu_\tau$, which presents a key challenge in the above program. So far, the evaluation of radiative corrections is based on chiral perturbation theory (ChPT)~\cite{Cirigliano:2001er,Cirigliano:2002pv} in combination with resonance models~\cite{Flores-Baez:2006yiq,Flores-Baez:2007vnd,Davier:2010fmf,Miranda:2020wdg,Davier:2023fpl,Castro:2024prg}, which makes robust uncertainty quantification difficult. Most importantly, we consider structure-dependent virtual corrections based on a dispersive representation of the pion form factor, which were previously shown to be sizable in the case of $e^+e^-\to\pi^+\pi^-$~\cite{Ignatov:2022iou,Colangelo:2022lzg}. In this context, we pay particular attention to the matching to ChPT, to be able to make the connection to short-distance (SD) contributions as well as lattice QCD~\cite{Bruno:2018ono}. We also revisit the calculation of real corrections, including the singular threshold behavior, and develop a method for a stable numerical evaluation of these corrections down to the two-pion threshold. Finally, we perform fits to the available $\tau$ spectral functions from Belle~\cite{Belle:2008xpe}, ALEPH~\cite{ALEPH:2005qgp,Davier:2013sfa}, CLEO~\cite{CLEO:1999dln}, and OPAL~\cite{OPAL:1998rrm}, to assess the consequences for the IB corrections in a $\tau$-based evaluation of the $2\pi$ channel, $a_\mu^\text{HVP, LO}[\pi\pi,\tau]$.

\emph{Formalism}---At first order in IB, the decay rate for the process $\tau\to\pi\pi\nu_\tau(\gamma)$ can be expressed in the form
\beq
\frac{d\Gamma[\tau\to\pi\pi\nu_\tau(\gamma)]}{ds}=S_\text{EW}^{\pi\pi}K_\Gamma(s)\beta_{\pi\pi^0}^3|f_+(s)|^2 G_\text{EM}(s),
\eeq
where $s$ is the invariant mass squared of the two-pion pair, the phase-space factor is defined as
\beq
\beta_{\pi\pi^0}=\lambda^{1/2}\Big(1,\frac{\mpi^2}{s},\frac{\mpii^2}{s}\Big),
\eeq
where $\mpi\equiv M_{\pi^\pm}$, $\lambda(a,b,c)=a^2+b^2+c^2-2(ab+ac+bc)$, 
and the normalization further involves
\beq
K_\Gamma(s)= \frac{\Gamma_e \vert V_{ud}\vert^2}{2 m_\tau^2} \bigg(1-\frac{s}{m_\tau^2}\bigg)^2\bigg(1+\frac{2s}{m_\tau^2}\bigg),
\eeq
with $\Gamma_e\equiv \Gamma[\tau\to e\nu_\tau\bar{\nu}_e]$, $\Br[\tau\to e\nu_\tau\bar{\nu}_e]=17.82(4)\%$~\cite{ParticleDataGroup:2024cfk,HeavyFlavorAveragingGroupHFLAV:2024ctg}, and $V_{ud}=0.97367(32)$~\cite{ParticleDataGroup:2024cfk}. We will also use the result of the global fit $\Br[\tau\to \pi\pi\nu_\tau]=25.49(9)\%$~\cite{ParticleDataGroup:2024cfk,HeavyFlavorAveragingGroupHFLAV:2024ctg}, including its correlation $-0.19$ with $\Br[\tau\to e\nu_\tau\bar{\nu}_e]$.

\begin{figure}[t]
 \centering
     \includegraphics[width=\linewidth]{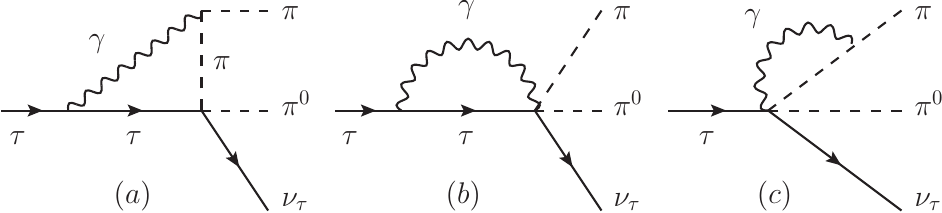}
     \caption{Virtual photonic corrections to $\tau\to\pi\pi\nu_\tau$ in ChPT. Solid lines denote leptons, dashed lines pions, and wiggly lines photons. Diagram $(a)$ refers to the chiral analog of the box diagram, while diagrams $(b)$ and $(c)$ originate from the five-particle vertex in ChPT. Not shown are self-energy diagrams and contact terms.}
     \label{fig:diagrams_ChPT}
\end{figure}
The SD contributions are subsumed into the electroweak (EW) correction factor
$S_\text{EW}^{\pi\pi}=1 + 2 \alpha/\pi\log (M_Z/m_\tau)    + \cdots=1.0233(3)(24)$~\cite{Sirlin:1981ie,Marciano:1985pd,Marciano:1988vm,Marciano:1993sh,Braaten:1990ef,Erler:2002mv,Davier:2002dy,Cirigliano:2023fnz}, where the uncertainty reflects the $\Order(\alpha/\pi)$ scheme ambiguity pointed out in Ref.~\cite{Aliberti:2025beg}. $f_+(s)$ denotes the pion form factor in the charged current, normalized as $f_+(0)=1$, and radiative corrections are expressed in terms of the electromagnetic (EM) correction factor
\beq
\label{G_EN_def}
 G_\text{EM}(s)=\frac{\int_{t_\text{min}(s)}^{t_\text{max}(s)}dt D(s,t)\big[1+2f_\text{loop}(s,t)+g_\text{rad}(s,t)\big]}{\int_{t_\text{min}(s)}^{t_\text{max}(s)}dt D(s,t)},
\eeq
which is the central object of this Letter. It is normalized in such a way that $G_\text{EM}(s)=1+\Order(\alpha)$, and amounts to averaging the amplitudes $f_\text{loop}(s,t)$ and $g_\text{rad}(s,t)$ for virtual and real contributions, respectively, over the entire phase space. The variable $t$ represents the invariant mass squared of the $\tau$ and the charged pion, with integration boundaries $t_{\text{min}/\text{max}}(s)$,
and the weight function reads
\beq
D(s,t)=\frac{m_\tau^2}{2}\big(m_\tau^2-s\big)+2\mpi^4-2t\big(m_\tau^2-s+2\mpi^2\big)+2t^2.
\eeq
In practice, we perform the calculation of $G_\text{EM}(s)$ in the isospin limit, as otherwise a large number of second-order IB effects would need to be considered, and restore the correct phase-space boundaries by mapping $[(\mpi+\mpii)^2,m_\tau^2]$ linearly onto $[4\mpi^2,m_\tau^2]$ in the end~\cite{Colangelo:2018jxw}.

\emph{Virtual corrections}---The virtual corrections in ChPT are shown in Fig.~\ref{fig:diagrams_ChPT}. Using the Lagrangians constructed in Refs.~\cite{Urech:1994hd,Knecht:1999ag}, the complete corrections at $\Order(e^2p^2)$ were derived in Refs.~\cite{Cirigliano:2001er,Cirigliano:2002pv}. In particular, adding self-energy and contact-term contributions, this leads to a chiral prediction $f_\text{loop}^\text{ChPT}(t)$ that is UV finite but IR divergent, with the latter divergences to be canceled by real radiation. The validity of the resulting prediction for $G_\text{EM}(s)$, however, is limited to the low-energy region, as the physics of hadronic resonances is only reproduced perturbatively or integrated out in the low-energy constants (LECs).

\begin{figure}[t]
 \centering
     \includegraphics[width=0.4\linewidth]{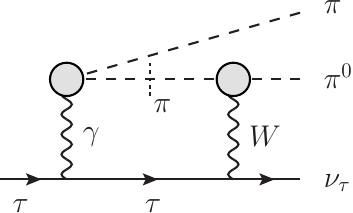}
     \caption{Box diagram in a dispersive approach. The gray blobs denote the pion form factor, in the neutral and charged channel, respectively. The short-dashed line indicates that the intermediate-state pion is taken on-shell.}
     \label{fig:diagrams_disp}
\end{figure}

This limitation can be overcome in a dispersive approach, in which case the diagrams in Fig.~\ref{fig:diagrams_ChPT} need to be reinterpreted as unitarity diagrams. From this perspective, since diagrams $(b)$ and $(c)$ do not generate analytic singularities in the Mandelstam variables,
only diagram $(a)$ needs to be considered further. Keeping in mind that the $\tau\nu_\tau \pi\pi^0$ vertex arises from integrating out the $W$ boson, this topology then takes the form shown in Fig.~\ref{fig:diagrams_disp}, and, accordingly, is referred to as ``box diagram.'' In general, its calculation requires knowledge of the full amplitude $\langle \pi\pi^0|j_\text{EM}^\mu j_W^\nu|0\rangle$ of EW and EM currents, which could be constructed dispersively in analogy to pion Compton scattering~\cite{Garcia-Martin:2010kyn,Hoferichter:2011wk,Moussallam:2013una,Danilkin:2018qfn,Hoferichter:2019nlq,Danilkin:2019opj}. In practice, by far the most dominant correction will arise from the pion-pole contribution indicated in Fig.~\ref{fig:diagrams_disp}. Following the strategy of Ref.~\cite{Colangelo:2022lzg}, the calculation can be reduced to standard one-loop Passarino--Veltman functions $B_0$, $C_0$, $D_0$~\cite{Passarino:1978jh} by inserting a dispersive representation of the pion form factor
\beq
\label{f_disp}
f_+(s)=\frac{1}{\pi}\int_{4\mpi^2}^\infty ds'\frac{\Im f_+(s')}{s'-s}
\eeq
and interpreting the Cauchy kernel as a propagator in the loop calculation. In this way, we arrive at a decomposition
\begin{align}
\label{f_disp_final}
f_\text{loop}^\text{disp}(s,t)&=\alpha \int_{4\mpi^2}^\infty ds'\int_{4\mpi^2}^\infty ds''\,\Im f_+(s')\Im f_+(s'')\notag\\
&\times\sum_{k\in\{B_0,C_0,D_0\}}{\mathcal M}_k(s,t,s',s''),
\end{align}
where the ${\mathcal M}_k$ represent a collection of loop functions and kinematic factors. In particular, using an unsubtracted dispersion relation~\eqref{f_disp} ensures that the result is manifestly UV finite. Moreover, we checked that IR singularities and the chiral logarithms agree with ChPT, by evaluating Eq.~\eqref{f_disp_final} in the limit of a narrow resonance. Numerically, the evaluation becomes complicated by a superficial singularity at $s'=s''$ in one of the IR-divergent $D_0$ functions~\cite{Beenakker:1988jr}, which disappears due to a factorization property at the border of the phase space---a cancellation that needs to be made manifest for a stable implementation. Based on the same factorization one can show explicitly that the resulting contribution to $G_\text{EM}(s)$ is finite at threshold, in agreement with the expectation from ChPT.
More details of our dispersive result for $G_\text{EM}(s)$ are given in Ref.~\cite{Colangelo:2025ivq}, which contains all the details concerning the analyses presented in this Letter.

The dispersive calculation of $f_\text{loop}(s,t)$ does not yet account for the ChPT diagrams besides Fig.~\ref{fig:diagrams_ChPT}$(a)$, but due to the absence of any nontrivial analytic structure in the remaining ones, the full result can be constructed by matching at $s=t=0$:
\beq
f_\text{loop}^\text{full}(s,t)=f_\text{loop}^\text{disp}(s,t) - f_\text{loop}^\text{disp}(0,0) + f_\text{loop}^\text{ChPT}(0,0),
\eeq
which, as an added benefit, suppresses the high-energy part of the integral in Eq.~\eqref{f_disp_final} and thus reduces the sensitivity to the asymptotic form of $\Im f_+(s)$.

Finally, to produce numerical results we need to choose a value for the chiral LECs $X_\ell$~\cite{Descotes-Genon:2005wrq}, on which the final correction depends linearly, i.e., we can express our results at some value $\bar{X}_\ell(\mu)$ and restore the general result by means of~\cite{Colangelo:2025ivq}
\beq
G_\text{EM}(s)\big|_{X_\ell(\mu)}
= G_\text{EM}(s) \big|_{X_\ell(\mu) = \bar{X}_\ell(\mu)} - e^2 \big[X_\ell(\mu) - \bar{X}_\ell(\mu)\big].
\eeq
We choose $\bar{X}_\ell(M_\rho)=14\times10^{-3}$~\cite{Ma:2021azh}, while emphasizing that this is the place at which the scheme ambiguity in $S_\text{EW}^{\pi\pi}$ resurfaces, so that only the product $S_\text{EW}^{\pi\pi}G_\text{EM}(s)$ becomes scale and scheme independent at the considered order. Work to reduce this scheme dependence is in progress~\cite{Cirigliano:2026ios}, using input from lattice QCD for the required nonperturbative matrix elements in the matching~\cite{Feng:2020zdc,Yoo:2023gln}.

\begin{figure}[t]
 \centering
     \includegraphics[width=0.9\linewidth]{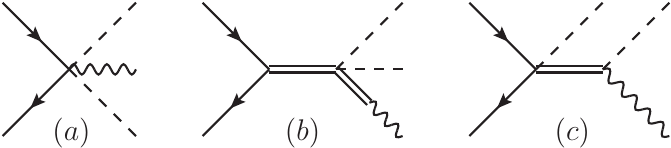}
     \caption{Sample diagrams for the real corrections from the WZW anomaly [diagram $(a)$] as well as the exchange of vector [diagram $(b)$] and axial-vector [diagram $(c)$] meson resonances (indicated by the double lines). Otherwise notation as in Figs.~\ref{fig:diagrams_ChPT} and~\ref{fig:diagrams_disp}.}
     \label{fig:diagrams_real}
\end{figure}

\emph{Real corrections}---For the real corrections we follow the general strategy from Refs.~\cite{Cirigliano:2001er,Cirigliano:2002pv}, including radiation off the $\tau$ and the charged pion, resonance contributions that saturate the LECs $L_9$, $L_{10}$~\cite{Gasser:1983yg,Gasser:1984gg}, and the prediction from the Wess--Zumino--Witten (WZW) anomaly~\cite{Wess:1971yu,Witten:1983tw}, see Fig.~\ref{fig:diagrams_real}.
The leading corrections from Low's theorem~\cite{Low:1958sn} can be given in an analytic form, while the remainder has to be calculated numerically. In contrast to the leading Low contribution, which only diverges logarithmically at threshold, it behaves as $G_\text{EM}(s) \propto 1/(s-4\mpi^2)$, in such a way that a stable numerical evaluation becomes challenging close to threshold. This issue can be circumvented by a suitable change of integration variables~\cite{Colangelo:2025ivq}---angular variables that yield stable expressions for amplitude times Jacobian---which allows us to evaluate both real and virtual contributions down to threshold in a robust manner. 

The main source of uncertainty originates from the resonance parameters in the diagrams of Fig.~\ref{fig:diagrams_real}$(b,c)$, which, in the notation of Refs.~\cite{Ecker:1989yg,Ecker:1988te}, can be expressed in terms of two vector couplings $F_V$, $G_V$, one axial-vector coupling $F_A$, and the mass parameters $M_V$, $M_A$. Identifying the latter ones with $M_\rho$ and $M_{a_1}$, this leaves two main strategies for the determination of the couplings: SD constraints~\cite{Ecker:1989yg}
\begin{align}
\label{RChT_SD}
F_V&=\sqrt{2}F_\pi\simeq 0.13\GeV,\quad
G_V=\frac{F_\pi}{\sqrt{2}}\simeq 0.065\GeV, \notag\\ 
F_A&=F_\pi\simeq 0.092\GeV,
\end{align}
and data-driven determinations
\beq
\label{RChT_data}
F_V\simeq 0.16\GeV,\quad G_V\simeq 0.065\GeV,\quad F_A\simeq 0.12\GeV,
\eeq
obtained from $\rho\to e^+e^-$, $\rho\to\pi\pi$, $K^*\to K\pi$~\cite{Colangelo:2021moe}, and $a_1\to\pi\gamma$~\cite{Zielinski:1984au}, respectively. The extraction of $F_A$ from $a_1\to\pi\gamma$ has been challenged in the literature mostly suggesting smaller values of $F_A$~\cite{Moussallam:1997xx,Knecht:2001xc,Cirigliano:2004ue,Garcia-Martin:2010kyn,Condo:1993xa,CLAS:2008zko}, but the situation is far from conclusive. Hoping to shed some light on the issue, we attempted yet another indirect determination, by assuming that the $a_1\to \pi\gamma$ decay proceeds via $a_1\to\rho\pi\to\pi\gamma$~\cite{Colangelo:2025ivq,Hoferichter:2017ftn,Zanke:2021wiq}, which gives values of $F_A=(0.07\ldots 0.13)\GeV$, depending on the assumptions for the $a_1$ parameters. In view of the substantial uncertainty especially in $F_A$, we define our central values as the average of fits performed with either Eq.~\eqref{RChT_SD} or Eq.~\eqref{RChT_data} as input and assign the difference between the results as $1\sigma$ uncertainty. The resulting sizable error band should therefore encompass a reasonably broad set of vector-meson parameters,
while including yet higher multiplets does not appear well motivated.

\begin{figure}[t]
 \centering
     \includegraphics[width=\linewidth]{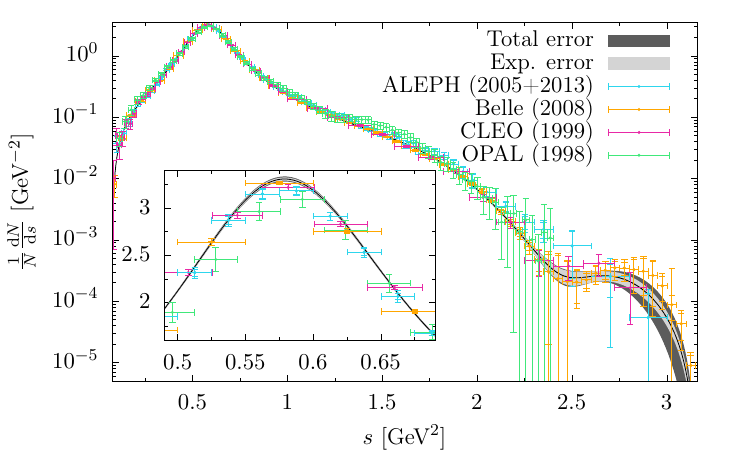}
     \caption{Global fit to the $\tau\to\pi\pi\nu_\tau$ spectrum, including the data sets from Belle~\cite{Belle:2008xpe}, ALEPH~\cite{ALEPH:2005qgp,Davier:2013sfa}, CLEO~\cite{CLEO:1999dln}, and OPAL~\cite{OPAL:1998rrm}.}
     \label{fig:fit}
\end{figure}

\emph{Fits to the $\tau$ spectral function}---For the numerical evaluation of Eq.~\eqref{f_disp_final} we need input for $\Im f_+(s)$, which should be determined in a self-consistent way with the data for the $\tau$ spectral function to which $G_\text{EM}(s)$ is applied. To achieve this, we employ an iterative procedure, starting from the Omn\`es function~\cite{Omnes:1958hv}
\beq
f_+(s)=\Omega^1_1(s)\equiv\exp\bigg\{\frac{s}{\pi}\int_{4\mpi^2}^\infty ds'\frac{\delta_1^1(s')}{s'(s'-s)}\bigg\},
\eeq
where $\delta^1_1$ denotes the $P$-wave $\pi\pi$ phase shift, and calculate a first approximation for $G_\text{EM}(s)$. We then use the resulting correction in a fit of $f_+(s)$ to the $\tau$ data, see End Matter for more details of the fit function (including a conformal polynomial with $N$ parameters to describe inelastic effects as well as explicit parameterizations of the $\rho'$, $\rho''$ resonances). We recalculate $G_\text{EM}(s)$ with the improved $f_+(s)$ obtained from the fit and iterate the procedure until it converges after a few steps.

We perform fits for $N=3,4,5$ to the Belle data only, to Belle+ALEPH, and to all data sets combined. This pattern is motivated by the fact that the Belle data set provides the most precise spectral function, necessary to resolve the detailed structure in the $\rho'$, $\rho''$ region. For the ALEPH data set also a full documentation of statistical and systematic covariance matrices is available, while for CLEO and OPAL systematic errors are either not provided or combined with the statistical ones. We take the global fit as our final result, but we still find it instructive to compare to fits to the more complete data sets only. Stable fits are obtained for $N=3,4$, with $p$-values ranging from a few percent for the Belle fits  to $\simeq 10^{-3}$ for the combined fits, signaling some tension in the data base. The fit quality improves significantly at $N=5$, to $p$-values of $\simeq 60\%$ and a few percent, respectively, but at the expense of clear signs of overfitting: the resonance parameters of the $\rho''$, which lies very close to the border of the phase space, need to be constrained, and the asymptotic behavior of $\Im f_+(s)$ outside the energy range covered by the decay region deteriorates significantly. For this reason, we choose the fits with $N=4$ as our central result, while including the variation to $N=3,5$ as a systematic error. The result of the global fit is shown in Fig.~\ref{fig:fit}.

\begin{figure}[t]
 \centering
     \includegraphics[width=\linewidth]{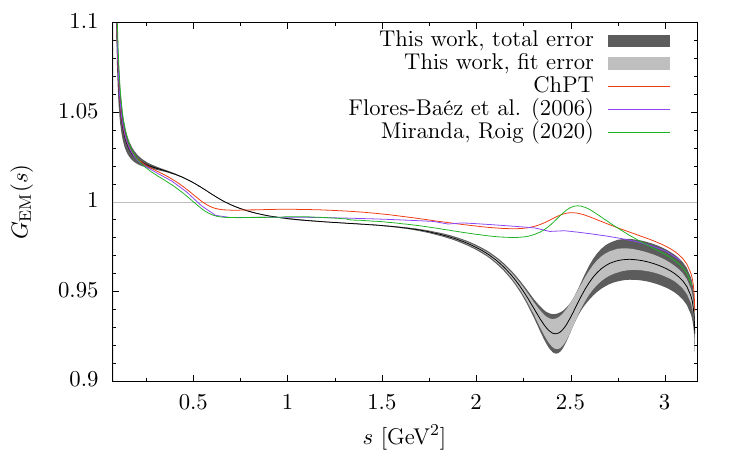}
     \caption{Final result for $G_\text{EM}(s)$ and comparison to previous work, Flores-Ba\'ez et al.\ (2006)~\cite{Flores-Baez:2006yiq}, Miranda, Roig (2020)~\cite{Miranda:2020wdg}, and using ChPT instead of dispersion relations for the box diagram (based on Refs.~\cite{Cirigliano:2001er,Cirigliano:2002pv}).}
     \label{fig:GEM}
\end{figure}

Based on these fits, we obtain the result for $G_\text{EM}(s)$ in Fig.~\ref{fig:GEM}. For small values of $s$ close to threshold, our result agrees with ChPT, while significant changes are observed in the vicinity of the $\rho$ resonance. These changes are
due to the structure-dependent virtual corrections from Fig.~\ref{fig:diagrams_disp}, and the corresponding increase in the $\rho$ region has immediate consequences for the $a_\mu$ integral.

\begin{table}[t]
	\centering
	\renewcommand{\arraystretch}{1.3}
	\begin{tabular}{lrrr}
	\toprule
	  & Belle & Belle+ALEPH & All\\\colrule
$G_\text{EM}^\text{Low}$ & $-2.292(15)(14)$ & $-2.279(13)(16)$ & $-2.267(13)(14)$\\
$G_\text{EM}^\text{rad}$ & $-5.21(3)(2)$ & $-5.20(3)(2)$ &$-5.19(3)(2)$ \\\colrule
$G_\text{EM}^\text{full}$ & $-5.44(3)(40)$ & $-5.43(3)(40)$ & $-5.41(3)(40)$\\
PS & $-7.74(4)(3)$ & $-7.73(4)(3)$ & $-7.74(4)(3)$\\
$S_\text{EW}^{\pi\pi}$ & $-12.180(57)(8)$ & $-12.177(57)(7)$ & $-12.166(56)(8)$\\
Sum & $-25.36(12)(44)$ & $-25.34(12)(44)$ & $-25.32(12)(44)$ \\\colrule
Full & $-24.84(12)(39)$ & $-24.82(12)(39)$ & $-24.80(12)(39)$\\\colrule 
$\tilde a_\mu$ & $510.1(2.4)(0.2)$ & $510.0(2.4)(0.2)$ & $509.5(2.4)(0.2)$\\
\botrule
	\renewcommand{\arraystretch}{1.0}
	\end{tabular}
	\caption{Summary of corrections to $a_\mu^\text{HVP, LO}[\pi\pi,\tau]$ in units of $10^{-10}$, see main text for the definition of the different rows. 
    All results are given for the fit to Belle, Belle+ALEPH, and all four data sets. The first (second) error refers to the experimental (theoretical) uncertainty, the latter excluding the scheme ambiguity in the SD contribution and higher intermediate states.}
	\label{tab:amu}
\end{table}

\emph{Consequences for $a_\mu$}---To quantify the impact on IB corrections to $a_\mu$, we use the master formula
\begin{align}
\Delta a_\mu^\text{HVP, LO}[\pi\pi,\tau]\big|_{r_\text{IB}}&=\bigg(\frac{\alpha m_\mu}{3\pi}\bigg)^2\int_{4\mpi^2}^{m_\tau^2}ds\frac{\hat K(s)}{4s^2}\notag\\
&\times\Big[r_\text{IB}(s)-1\Big]v_\tau(s),\\
v_\tau(s)&=S_\text{EW}^{\pi\pi}\big[\beta_{\pi\pi^0}\big]^3\big|f_+(s)\big|^2G_\text{EM}(s),\notag
\end{align}
where $r_\text{IB}(s)=\big\{1/G_\text{EM}(s), (\beta_{\pi\pi}/\beta_{\pi\pi^0})^3,1/S_\text{EW}^{\pi\pi}\big\}$ for the three $\tau$-specific IB contributions---$G_\text{EM}(s)$, phase space (PS), and SD---in a linearized form, and $\hat K(s)$ is the standard $a_\mu$ kernel function~\cite{Bouchiat:1961lbg,Brodsky:1967sr}. Our results for the three fit variants are summarized in Table~\ref{tab:amu}, where we also show the breakdown for the leading Low contribution, $G_\text{EM}^\text{Low}(s)$, and the full radiation off $\tau$ and $\pi$ without resonance or WZW effects, $G_\text{EM}^\text{rad}(s)$ (the latter are included in $G_\text{EM}^\text{full}(s)$). Moreover, we give the sum of the three contributions, the full correction without linearization, as well as a quantity $\tilde a_\mu$ defined by $r_\text{IB}(s)=1+(\beta_{\pi\pi}/\beta_{\pi\pi^0})^3/[G_\text{EM}(s)S_\text{EW}^{\pi\pi}]$, quantifying the $a_\mu$ integral after the $\tau$-specific IB corrections have been applied, but not IB corrections in the matrix element nor in $e^+e^-$ added. The main conclusions from Table~\ref{tab:amu} are as follows: first, the differences among the fit variants are very small, owing to the small overall size of the IB corrections. Second, at the relevant level of precision the linearization of IB corrections is not justified, as $\Order(e^4)$ terms enhanced by the threshold singularity of $G_\text{EM}(s)$, see Fig.~\ref{fig:GEM}, are sizable, emphasizing the importance of a stable numerical implementation down to threshold.

\begin{table}[t]
	\centering
	\renewcommand{\arraystretch}{1.3}
	\begin{tabular}{lrrrr}
	\toprule
& Ref.~\cite{Davier:2023fpl} & Ref.~\cite{Castro:2024prg} & Ref.~\cite{Aliberti:2025beg} & This Letter\\\colrule
PS & $-7.88$ & $-7.52$ & $-7.7(2)$ & $-7.74(5)$\\
$S_\text{EW}^{\pi\pi}$ & $-12.21(15)$ & $-12.16(15)$ & $-12.2(1.3)$ & $-12.2(1.3)$\\
$G_\text{EM}^\text{full}$ & $-1.92(90)$  & $(-1.67)^{+0.60}_{-1.39}$ & $-2.0(1.4)$ & $-5.4(5)$ \\
Sum & $-22.01(91)$ & $(-21.35)^{+0.62}_{-1.40}$ & $-21.9(1.9)$ & $-25.3(1.4)$ \\\colrule
Full & -- & -- & -- & $-24.8(1.4)$\\\colrule
$\tilde a_\mu$ & $510.3(3.0)$ & $510.9^{+2.9}_{-3.1}$ & $510.3(3.4)$ &$509.5(2.7)$\\
\botrule
	\renewcommand{\arraystretch}{1.0}
	\end{tabular}
	\caption{Comparison to previous work (notation as in Table~\ref{tab:amu}). The SD uncertainty from Ref.~\cite{Aliberti:2025beg} has been added to our results from Table~\ref{tab:amu}, as has an additional uncertainty of $0.3$ units to capture higher intermediate states in the virtual contribution to $G_\text{EM}(s)$~\cite{Colangelo:2025ivq}.}
	\label{tab:amu_comp}
\end{table}

Our results are compared to previous work in Table~\ref{tab:amu_comp}. For PS and SD corrections there is reasonable agreement, but the $G_\text{EM}(s)$ contribution increases in magnitude, as a consequence of the structure-dependent virtual diagrams and local terms in the matching to ChPT. Part of this shift is canceled when including the threshold-enhanced $\Order(e^4)$ terms, so that the net effect
\beq
\Delta a_\mu^\text{HVP, LO}[\pi\pi,\tau]\big|_\text{Full}=-24.8(0.1)_\text{exp}(0.5)_\text{th}(1.3)_\text{SD},
\eeq
changes by about three units (in $10^{-10}$). While the overall uncertainty is now dominated by the scheme dependence in the SD contribution, addressed in Ref.~\cite{Cirigliano:2026ios}, the remaining uncertainty due to radiative corrections mainly reflects the role of higher intermediate states in the real and virtual contributions. 
Overall, the uncertainty due to $G_\text{EM}(s)$ has been reduced by almost a factor of three over the previously assigned error in Ref.~\cite{Aliberti:2025beg}, with a shift in central value by about $2.5\sigma$.

Besides improving the radiative corrections parameterized by $G_\text{EM}(s)$, our work also strongly motivates increased efforts in new measurements of the $\tau\to\pi\pi\nu_\tau$ spectral function, as possible at Belle II~\cite{Belle-II:2018jsg}. Indeed, our dispersive fits to the spectrum reveal that some tensions among the currently available data sets do exist, and at the same time we observe differences to previous evaluations, as visible in the context of $\tilde a_\mu$ in Table~\ref{tab:amu_comp}. Part of the difference might originate from the changes in $G_\text{EM}(s)$~\cite{data}, but we also find that the constraints imposed by analyticity and unitarity result in a moderate tension between the low-energy part of the spectrum and the $\rho$ region, which tends to increase the integral for small values of $s$. An improved measurement of the $\tau\to\pi\pi\nu_\tau$ spectral function, profiting from the statistics available at Belle II, would therefore be extremely valuable.

 \begin{acknowledgments}
 \emph{Acknowledgments}---We thank  Hisaki Hayashii, Bogdan Malaescu,  Lucas Mansur, Sven Menke, and Zhiqing Zhang for correspondence on the $\tau\to\pi\pi\nu_\tau$ data sets. Further, we express our gratitude to Vincenzo Cirigliano, Christoph Greub, Pablo Roig, and Peter Stoffer for helpful discussions.
Financial support by the Swiss National Science Foundation (Project Nos.\ 200020\_200553 and TMCG-2\_213690) and the Albert Einstein Center for Fundamental Physics is gratefully acknowledged.
\end{acknowledgments}

\section{End Matter}

\emph{Dispersive fit function}---For the fit we use a form that combines the dispersive representation developed for $e^+e^-\to\pi^+\pi^-$~\cite{Colangelo:2018mtw,Colangelo:2022prz} with explicit resonance contributions for $\rho'$, $\rho''$~\cite{Hoferichter:2014vra,Hoferichter:2019mqg,Hoferichter:2023bjm,Hoferichter:2025lcz}:
\beq
f_+(s)=\bigg[1+G_\text{in}^N(s)+\sum_{V=\rho',\rho''}c_V{\mathcal A}_V(s)\bigg]\Omega^1_1(s),
\eeq
where the phase shift is parameterized as the solution of the Roy equations~\cite{Roy:1971tc,Ananthanarayan:2000ht,Caprini:2011ky} in terms of the phase-shift values at $s_0=(0.8\GeV)^2$, $s_1=(1.15\GeV)^2$, $G_\text{in}^N(s)$ denotes a conformal polynomial with threshold $s_\text{thr}=(\mpi+M_\omega)^2$, and the explicit form for ${\mathcal A}_V$ reads
\begin{align}
{\mathcal A}_V(s)&=\frac{s}{\pi}\int_{s_\text{thr}}^\infty ds'\frac{\Im {\mathcal A}_V(s')}{s'(s'-s)},\notag\\
\Im {\mathcal A}_V(s)&=\Im\frac{1}{M_V^2-s-i\sqrt{s}\Gamma_V(s)},
\end{align}
where the energy dependence of the width $\Gamma_V(s)$ is constructed from the $\pi\omega$ phase space. In this way, we arrive at two free parameters in $\Omega^1_1(s)$, three for each resonance, plus the $N-2$ free parameters in $G_\text{in}^N(s)$. Here, we use one parameter to ensure $P$-wave behavior of $\Im f_+(s)$, and a second parameter to enforce the sum rule
\beq
1=\frac{1}{\pi}\int_{4\mpi^2}^\infty ds'\frac{\Im f_+(s')}{s'},
\eeq
which, in the unsubtracted form~\eqref{f_disp}, is mandatory to respect charge conservation. We also tried fit variants in which $\rho'$, $\rho''$ are directly implemented via the conformal variable~\cite{Kirk:2024oyl}, but observed worse performance especially for the $\rho''$ due to its location very close to the border of the phase space. Finally, for the fit at each step we employ yet another iteration to avoid D'Agostini bias~\cite{DAgostini:1993arp,Ball:2009qv}, and account for the finite bin size via a weighted average~\cite{Colangelo:2018mtw}.

\bibliography{amu}

\end{document}